\documentclass[useAMS,usenatbib]{mn2e}
\pdfoutput=1
\usepackage{aas_macros}
\usepackage{epsfig}
\usepackage{amsmath}
\usepackage{amssymb}
\usepackage{leftidx}
\usepackage{natbib}
\usepackage{xcolor}
\usepackage{hyperref}
\hypersetup{colorlinks=true,citecolor=blue,linkcolor=violet}%,pdfborder={0 0 0},}
\def\beq{\begin{equation}}
\def\eeq{\end{equation}}
\def\ba{\begin{eqnarray}}
\def\ea{\end{eqnarray}}
\def\go{\mathrel{\raise.3ex\hbox{$>$}\mkern-14mu
             \lower0.6ex\hbox{$\sim$}}}
\def\lo{\mathrel{\raise.3ex\hbox{$<$}\mkern-14mu
             \lower0.6ex\hbox{$\sim$}}}
\def\bxi{{\mbox{\boldmath $\xi$}}}

\begin{document}

\title[Resonance Locking in KIC 8164262]{Accelerated Tidal Circularization Via Resonance Locking in KIC 8164262}
\author[Fuller et al.]
{Jim Fuller$^{1,2}$\thanks{Email:
jfuller@caltech.edu},
Kelly Hambleton$^{3}$,
Avi Shporer$^{4}$,
Howard Isaacson$^{5}$, 
\and Susan Thompson$^{6}$ \\
\\ $^1$ Kavli Institute for Theoretical Physics, Kohn Hall, University of California, Santa Barbara, CA 93106, USA
\\ $^2$ TAPIR, Walter Burke Institute for Theoretical Physics, California Institute of Technology, Pasadena, CA 91125, USA
\\ $^3$ Department of Astrophysics and Planetary Science, Villanova University, 800 East Lancaster Avenue, Villanova, PA 19085, USA
\\ $^4$ Division of Geological and Planetary Sciences, California Institute of Technology, Pasadena, CA 91125
\\ $^5$ Astronomy Department, University of California, Berkeley, CA 94720, USA
\\ $^6$ SETI Institute/NASA Ames Research Center, Moffett Field, CA 94035, USA
}

\label{firstpage}
\maketitle
%%%%%%%%%%%%%%%%%%%%%%%%%%%%%%%%%%%%%%%%%%%%%%%%%%%%%%%%%%%%%%%%%%%%
\begin{abstract}

Tidal dissipation in binary star and planetary systems is poorly understood. Fortunately, eccentric binaries known as heartbeat stars often exhibit tidally excited oscillations, providing observable diagnostics of  tidal circularization mechanisms and timescales. We apply tidal theories to observations of the heartbeat star KIC 8164262, which contains an F-type primary in a very eccentric orbit that exhibits a prominent tidally excited oscillation. We demonstrate that the prominent oscillation is unlikely to result from a chance resonance between tidal forcing and a stellar oscillation mode. However, the oscillation has a frequency and amplitude consistent with the prediction of resonance locking, a mechanism in which coupled stellar and orbital evolution maintain a stable resonance between tidal forcing and a stellar oscillation mode. The resonantly excited mode produces efficient tidal dissipation (corresponding to an effective tidal quality factor $Q \sim 5 \times 10^4$), such that tidal orbital decay/circularization proceeds on a stellar evolution time scale.

\end{abstract}


\section{Introduction}
\label{intro}

The mechanisms underlying tidal energy dissipation in stellar and gaseous planetary interiors remain uncertain despite decades of research. Observationally, tidal orbital evolution is challenging to measure because it typically proceeds on very long timescales. Theoretically, tidal dissipation is difficult to calculate from first principles because it is produced by weak friction effects that depend on details of the stellar structure, convective turbulence, non-linear mode coupling, and other complex hydrodynamical processes.

Heartbeat stars present a new opportunity to constrain tidal dissipation processes through observations of tidally excited oscillations (TEOs). These eccentric binary stars experience tidal distortion near periastron that produces ``heartbeat" signals in high precision light curves, and they have been studied in a number of recent works \citep{welsh:11,thompson:12,hambleton:13,beck:14,schmid:15,smullen:15,hambleton:16,kirk:16,shporer:16,dimitrov:17,guo:17}. A fraction of heartbeat stars also exhibit TEOs that can be recognized because they occur at {\it exact} integer multiples of the orbital frequency \citep{kumar:95,fullerkoi54:12,burkart:12}. The TEOs are produced by tidally forced stellar oscillation modes, in most cases by gravity modes (g modes). Given sufficiently accurate stellar properties, one can identify the g modes responsible for the TEOs. From observed mode amplitudes, one can then calculate mode energies, damping rates, tidal energy dissipation rates, and circularization/synchronization time scales.

\begin{figure*}
\begin{center}
\includegraphics[scale=0.58]{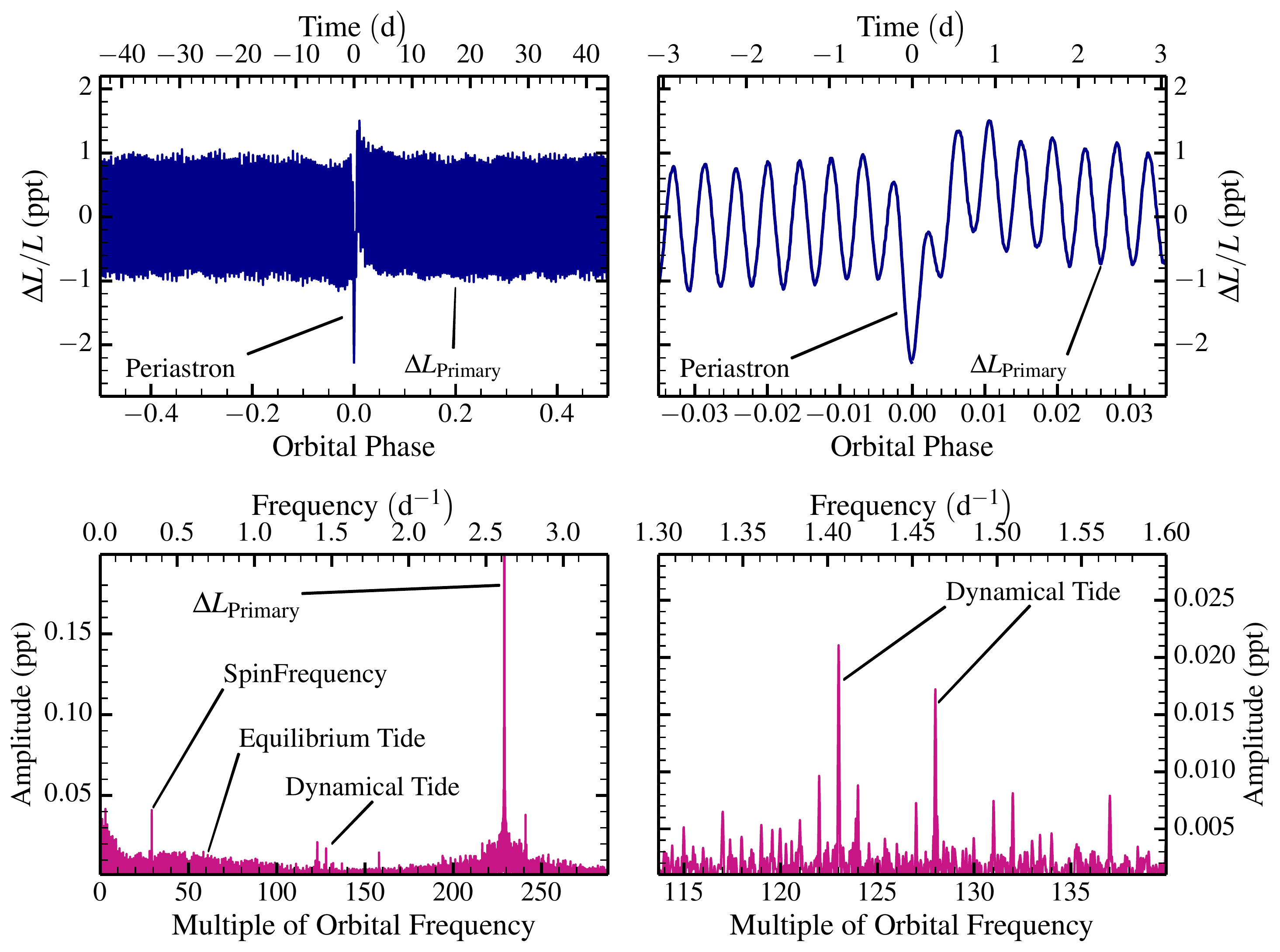}
\end{center} 
\caption{ \label{KIC81lc}  {\bf Top Left:} Phased lightcurve of KIC 8164262, in units of parts per thousand, with periastron near phase zero. The sharp variation near periastron is the ``heartbeat" signal produced by the equilibrium tidal distortion, reflection, and Doppler boosting. The rapid variability away from periastron is due to the prominent tidally excited oscillation at 229 times the orbital frequency. {\bf Top Right:} Same as top left panel, zoomed in near periastron. {\bf Bottom Left:} Fourier transform of the lightcurve of KIC 8164262. The large amplitude spike at $f \simeq 2.6\,{\rm d}^{-1}$ is produced by the dominant tidally excited oscillation and extends above the scale of the plot to 1 ppt. The fuzz at lower amplitudes is composed of peaks at integer multiples of the orbital frequency. {\bf Bottom Right:} Same as bottom left panel, zoomed in to frequencies near $1.5 \, {\rm d}^{-1}$. Peaks at orbital harmonics are produced by tidally excited oscillations (dynamical tide).}
%The series of evenly spaced peaks are located at integer harmonics of the orbital frequency, and are generated primarily by the heartbeat signal in the light curve. The peak at $f\simeq 0.33 \, {\rm d}^{-1}$ is not an orbital harmonic, and likely represents the spin frequency of the primary star (H15). }
\end{figure*}

In this paper, we compare detailed tidal theory from a companion paper \citep{fullerhb:17}, with observations of the heartbeat star KIC 8164262 (hereafter K81) analyzed in another companion paper \citep{hambleton:17}. This heartbeat star contains an F-type primary star in a highly eccentric ($ e\simeq 0.89$), long period ($P=87\,{\rm d}$) orbit with a low mass ($M'\approx 0.36 M_\odot$) companion. The F-type primary has $M =1.7 \pm 0.1 \, M_\odot$, $R = 2.4 \pm 0.1 \, R_\odot$, $T_{\rm eff} = 6900 \pm 100 \, {\rm K}$, and is nearing the end of its main sequence lifetime.  Most importantly, K81 exhibits a large amplitude (relative flux variability of $\Delta L/L \sim 10^{-3}$) TEO at exactly 229 times the orbital frequency. A light curve and power spectrum is shown in Figure \ref{KIC81lc}. 

TEOs can be excited to large amplitude if a stellar g mode frequency happens to be nearly equal to a tidal forcing frequency (i.e., a multiple of the orbital frequency). We demonstrate that the prominent TEO in K81 requires a resonance so finely tuned that it is unlikely to occur by chance. Instead, the oscillation presents a compelling case for resonance locking (\citealt{witte:99,witte:01,fullerkoi54:12,burkart:14}), where tidal orbital decay naturally maintains a finely tuned resonance with a g mode, resulting in a large amplitude TEO. We show that the prominent TEO in K81 has a frequency and amplitude consistent with this hypothesis. The resonance locking results in tidal orbital decay/circularization that occurs on a stellar evolution time scale, much faster than expected for this system.

\section{Tidal Models of  KIC 8164262}
\label{kic81}

TEOs are produced by tidally forced stellar oscillation modes. Computing expectations for TEO frequencies, amplitudes, and phases is discussed in detail in \cite{fullerhb:17}. In many cases such as in K81, only the highest amplitude TEOs are detectable, and these TEOs are produced by near-resonances between stellar g mode frequencies and multiples of the orbital frequency. 

\begin{figure}
\begin{center}
\includegraphics[scale=0.43]{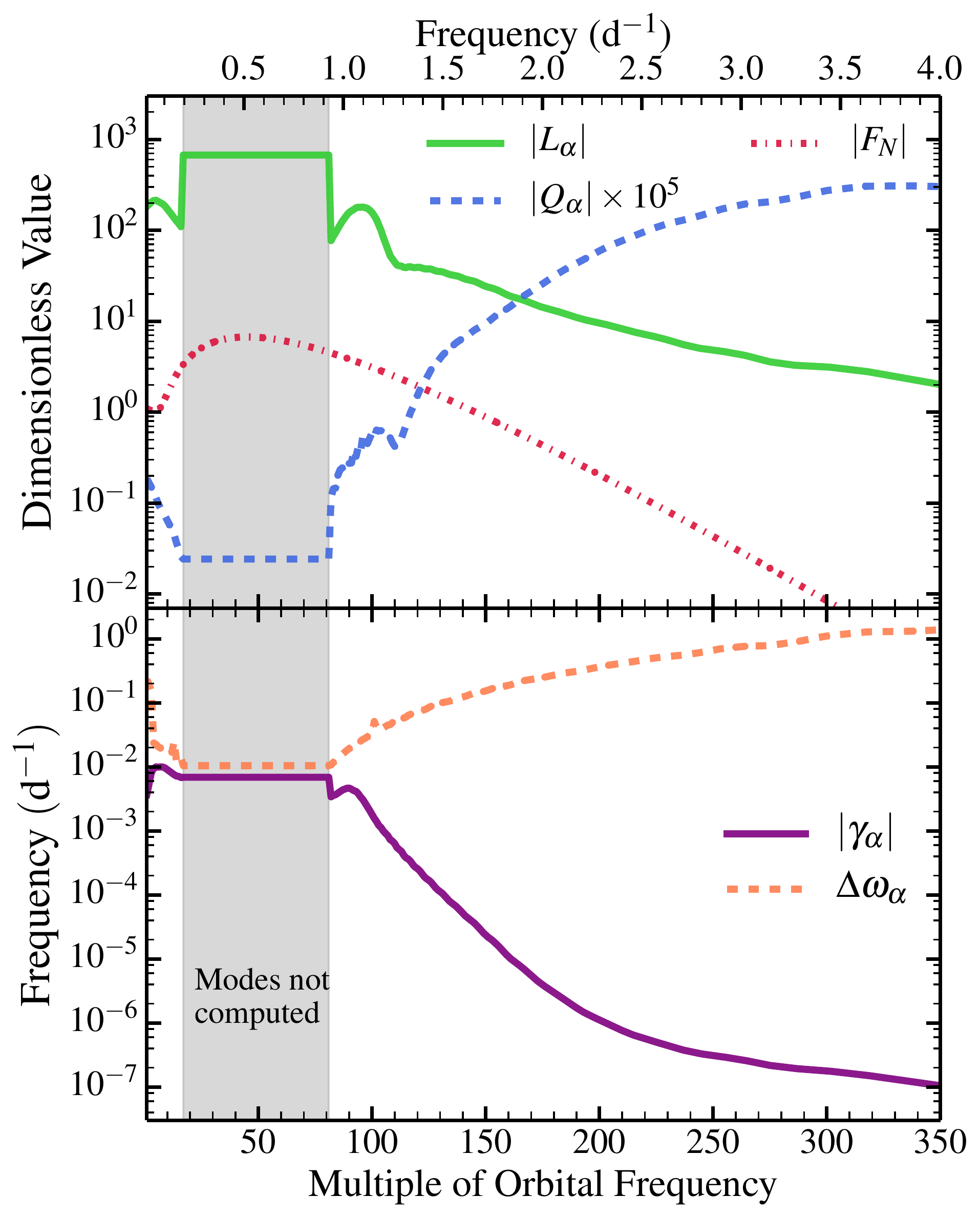}
\end{center} 
\caption{ \label{modeplot4}  {\bf Top:} Properties of normalized $l=m=2$ g modes in our model of the primary star of KIC 8164262 (see text). All values have been calculated at multiples $N$ of the orbtital frequency by interpolating between neighboring g modes. Modes within the grey shaded region have very large radial orders and have not been calculated. The top axis shows the corresponding observed oscillation frequency $f$. {\bf Bottom:} Mode damping rate $\gamma_\alpha$, and mode frequency separation $\Delta \omega_\alpha$. Note that most mode damping times are longer than $10^3$ days so that tidally excited oscillations maintain a nearly constant amplitude throughout each orbit.}
\end{figure}

When a single resonant oscillation mode (labeled by $\alpha$) dominates the tidal response at a multiple $N$ of the orbital frequency, it produces a sinusoidal luminosity fluctuation of form
\beq
\label{dlumres}
\frac{\Delta L_{N}}{L} \simeq A_N \sin(N \Omega t + \Delta_N) \, 
\eeq
where $\Omega=2 \pi/P$ is the angular orbital frequency. The amplitude is
\begin{align}
\label{Aalpha}
A_N &= \epsilon V_{lm} X_{Nm}\big| Q_{\alpha} L_{\alpha} \big| \frac{\omega_{Nm}}{\sqrt{(\omega_{\alpha} - \omega_{Nm})^2 + \gamma_{\alpha}^2 }} \, .
\end{align}
The pulsation phase $\Delta_N$ also contains useful information \citep{oleary:14}, but the longitude of periastron measurement in K81 from \cite{hambleton:17} was not precise enough to measure the phases of its TEOs. Each term in equation \ref{Aalpha} is defined in \cite{fullerhb:17} and can be calculated relatively easily. The tidal forcing amplitude $\epsilon$ has a value of $\epsilon \simeq 10^{-6}$ for K81. $V_{lm}$ describes the visibility of the modes based on viewing angle, and is typically of order unity. Other quantities are a function of frequency and are plotted in Figure \ref{modeplot4}. The Hansen coefficient $X_{Nm}$ is the strength of tidal forcing at   each forcing frequency $N \Omega$ and peaks near $N\sim40$ for $m=2$ for K81 (see Figure \ref{modeplot4}). It has a long tail to higher frequencies (due to the high eccentricity) allowing oscillations up to $N \sim 300$ to be observable.

The other terms in equation \ref{Aalpha} are properties of stellar oscillation modes. To calculate them, we construct a stellar model with parameters nearly equal to those found by \cite{hambleton:17} for the primary star, using the stellar evolution code MESA \citep{paxton:11,paxton:13,paxton:15}. Our model has the same stellar/orbital parameters quoted above, but with a $1 \sigma$ smaller stellar radius of $2.3 \, R_\odot$ that better matched the observed TEOs. We calculate the non-adiabatic stellar oscillation modes of our model using the GYRE oscillation code \citep{townsend:13}. MESA/GYRE inlists and discussion of the process are provided in \cite{fullerhb:17}. Next, we calculate tidal overlap integrals $Q_\alpha$ that measure gravitational coupling between oscillation modes and the tidal potential, and bolometric luminosity perturbations $L_\alpha$ produced by normalized modes at the stellar surface. Figure \ref{modeplot4} shows $Q_\alpha$ and $L_\alpha$ for $l=|m|=2$ modes (we show below that most of the observed oscillations are likely generated by quadrupolar, $m=1$ and $m=2$ oscillation modes). Although the value of $Q_\alpha$ peaks for high frequency (low radial order) g modes, the value of $L_\alpha$ peaks for modes with $N \sim 100$. Figure \ref{modeplot4} also shows mode damping rates $\gamma_\alpha$ and frequency spacings $\Delta \omega_\alpha$. Damping rates are much larger for low frequency (high radial order) modes, where the mode spectrum is very dense. Modes behave like traveling waves when $\gamma_\alpha \sim \Delta \omega_\alpha$, which occurs at frequencies $N \lesssim 100$.

%The peak in $L_\alpha$ is as follows. Higher frequency modes become evanescent below the photosphere (because their frequency is above the local Lamb frequency) and thus have smaller surface perturbations. Lower frequency modes, on the other hand, are strongly damped via non-adiabatic effects. Consequently, they behave like outward traveling waves, and their amplitude is attenuated as they propagate toward the photosphere, once again resulting in small surface perturbations. The peak in $L_\alpha$ occurs at frequencies corresponding to a transition between standing oscillation modes and traveling waves, and entails that oscillations near this frequency are especially likely to be observed. 

The last term in equation \ref{Aalpha} is the resonant deturning for with $\omega_\alpha \simeq \omega_{Nm}$. The detuning factor can be very large for nearly resonant modes but is very sensitive to the degree of detuning. Because of this extreme sensitivity, it is difficult to reliably calculate this quantity. However, we can still predict mode amplitudes as a function of frequency in a stasticical fashion, as outlined in \cite{fullerhb:17}. The median luminosity fluctuation is
\beq
\label{dlummean}
A_{N,{\rm med}} \simeq \bigg| 4 \mathcal{L}_{N} \frac{\omega_{Nm}}{\Delta \omega_\alpha} \bigg|.
\eeq
while the maximum possible fluctuation amplitude is
\beq
\label{dlummax}
A_{N,{\rm max}} \simeq \bigg| \mathcal{L}_{N} \frac{\omega_{Nm}}{\gamma_\alpha} \bigg| \, ,
\eeq
where $\mathcal{L}_N = \epsilon V_{lm} X_{Nm}\big| Q_{\alpha} L_{\alpha} \big|$.

\begin{figure*}
\begin{center}
\includegraphics[scale=0.53]{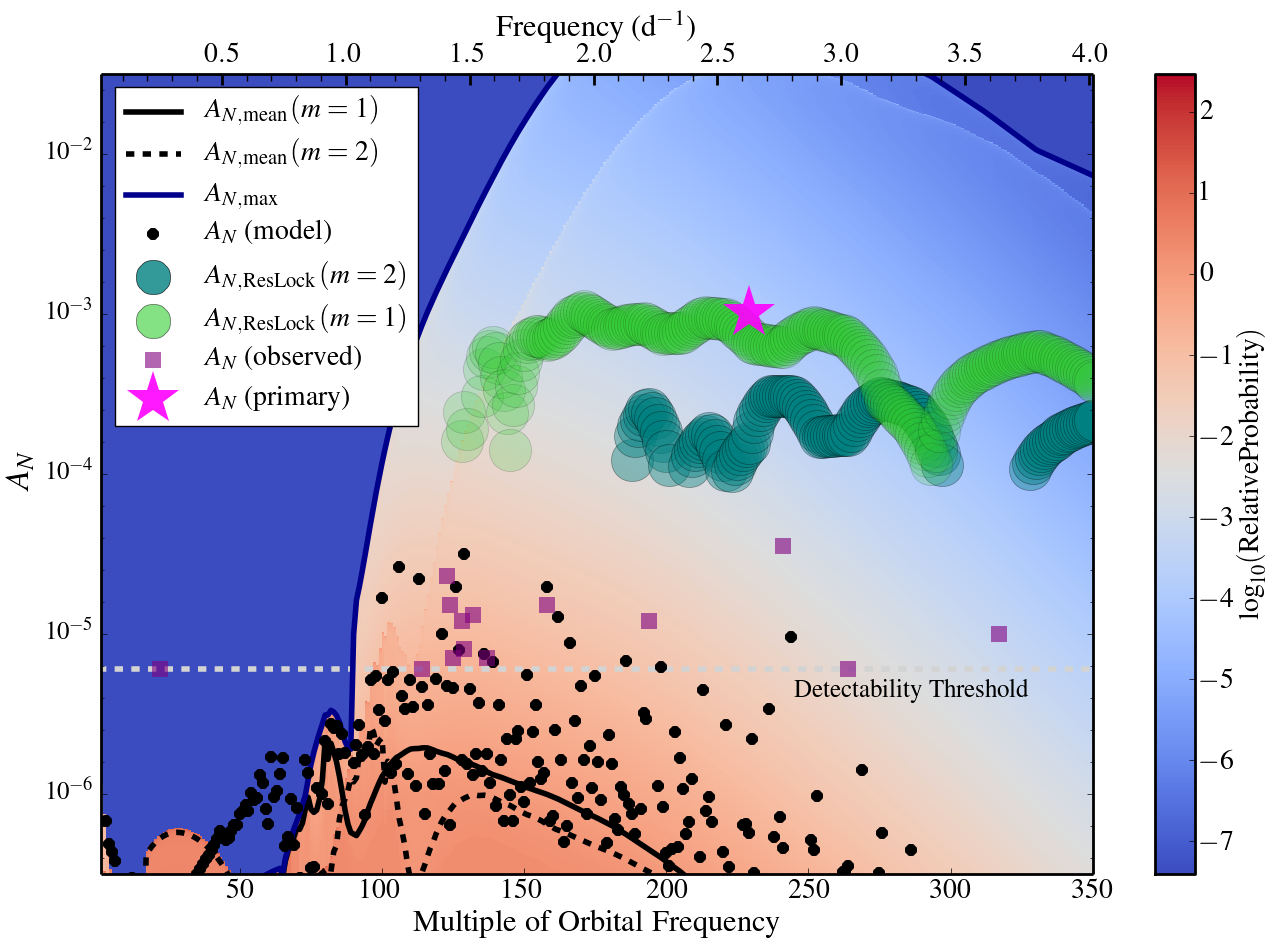}
\end{center} 
\caption{ \label{KIC81plot} TEOs in the KIC 8164262 system. Observed TEOs are shown as purple squares, with the prominent TEO denoted by a magenta star. Black dots are TEOs from our representative model. Black lines are the median amplitudes of the $m=1$ and $m=2$ TEOs in our modle, while the blue line is the maximum possible amplitude of $m=1$ or $m=2$ TEOs in our model. Background color denotes the probability density for $m=1$ and $m=2$ TEOs, which are more likely to exist in red regions of the plot. Note the primary TEO lies in a very unlikely region of parameter space. Blue-green and light green circles are expected amplitudes for resonantly locked $m=2$ and $m=1$ modes. }
\end{figure*}

With the K81 parameters from \cite{hambleton:17} and the oscillation modes computed above, we can compute expected frequencies and amplitudes of TEOs in K81. Figure \ref{KIC81plot} shows both the observed TEOs and our model results. We plot the median luminosity fluctuation at each orbital harmonic $N$ (equation \ref{dlummean}) and the maximum luminosity fluctuation (equation \ref{dlummax}) for $|m|=2$ modes. In addition, we plot the luminosity fluctuations at each harmonic for our representative stellar model. Finally, the background color indicates the probability density $d \log A_N/d N$ of observing an $m=2$ TEO in that part of the plot. Most of the observed TEOs in K81 are likely to be produced by quadrupolar $m=2$ and $m=1$ modes, with amplitudes above the median amplitude of equation \ref{dlummean} due to chance resonances between mode frequencies and forcing frequencies. In our model, $m=0$ modes are less visible due to the viewing angle relative to the stellar spin axis.

%The cluster of observed TEOs near $N\sim 130$ can be explained via the peak in $L_\alpha$ (Figure \ref{modeplot4}) that make it more likely to observe modes near these frequencies. The observed TEOs at larger frequencies are mostly due to chance resonances, which are unlikely at each harmonic $N$ but have many opportunities to occur because there are many harmonics at which TEOs can be excited to visible amplitudes. 

\cite{hambleton:17} found evidence for spin-orbit misalignment in K81, measuring an orbital inclination of $i_o = 65 \pm 1^\circ$ and a spin inclination of $i_s = 35 \pm 3^\circ$, implying a minimum obliquity of $30^\circ$. Such spin-orbit misalignment allows for excitation of quadrupolar $m=1$ modes. The model shown in Figure \ref{KIC81plot} has $i_s = 40^\circ$ and an actual (non-projected) obliquity of $\beta = 30^\circ$. We find models with lower values of $i_s$ or higher obliquity provide a worse fit to the data because they produce an excess of low-frequency ($N \lesssim 110$) $m=1,0$ TEOs that are not observed. 

Another modeling feature that improves agreement is the use of small amounts of convective core overshoot and diffusive mixing. Our model uses $f_{\rm ov} = 0.01$ and $D_{\rm mix} = 1 \, {\rm cm}^2/{\rm s}$, on the low side of values inferred from aseteroseismology \citep{moravveji:15,moravveji:16,deheuvels:16} which are typically in the range $0.01 \lesssim f_{\rm ov} \lesssim 0.03$ and $1 \, {\rm cm}^2/{\rm s}  \lesssim D_{\rm mix} \lesssim 100 \, {\rm cm}^2/{\rm s}$. Larger overshoot tends to overpredict TEO amplitudes. Our models use the Schwarzschild criterion for convective boundaries, appropriate for main sequence stars of this mass \citep{deheuvels:16,moore:16}. In principle, TEO amplitudes could constrain uncertainties such as convective boundary mixing, but this would require a more thorough examination of models over a multidimensional parameter space.

%Our favored model reduces contributions from $m=0$ modes because the visibility $V_{lm}$ goes to zero at $i_s \simeq 55^\circ$, and $m=1$ modes are sub-dominant because they are more easily excited for large obliquities (because $X_{Nm} \propto \sin \beta$).

\section{Resonance Locking in KIC 8164262}

The primary TEO at $N=229$ is not easily explained as a chance resonance. Figure \ref{KIC81plot} shows that this TEO lies in a region of parameter space unlikely to contain TEOs, in contrast to TEOs at lower frequencies and amplitudes. A very close resonance is required to produce the primary TEO, which is possible but unlikely.

\begin{figure}
\begin{center}
\includegraphics[scale=0.29]{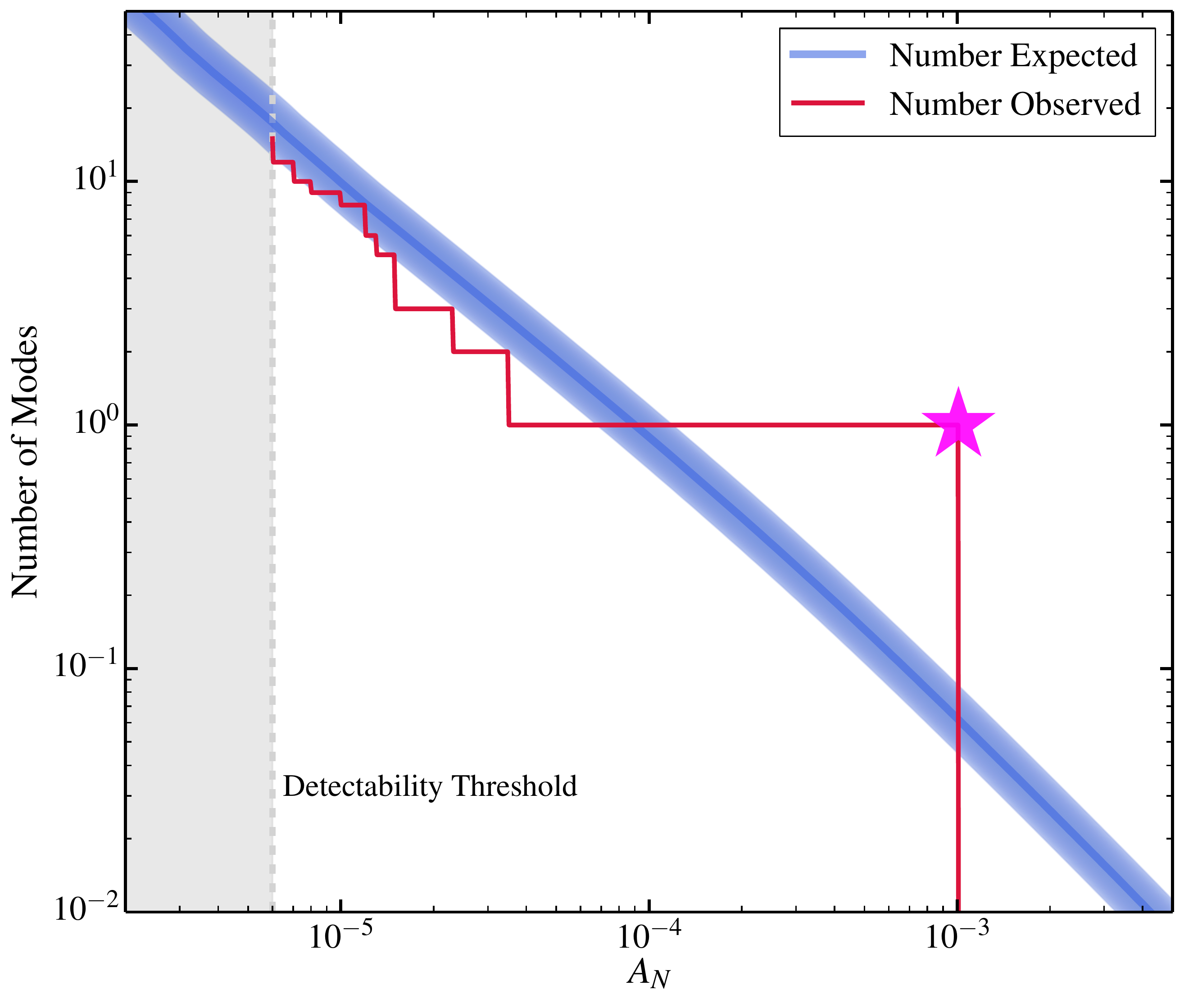}
\end{center} 
\caption{ \label{KIC81numex} Cumulative distribution of number of TEOs above an amplitude $A_N$ as a function of $A_N$. We plot both observed numbers and expected numbers from our representative model. At low amplitudes, the observed and expected distributions overlap, indicating the model adequately explains most of the low-amplitude TEOs. The primary TEO at $A_N = 10^{-3}$ (magenta star) is unexpected from the model and is unlikely to occur by a chance resonance (the probability is $\approx 5 \%$), and instead could be the result of resonance locking.}
\end{figure}

To quantify this statement, we compute the cumulative distribution of the expected number of TEOs to exist above a given amplitude $\Delta L/L$, and compare with the distribution of observed TEO amplitudes. This calculation is described in \cite{fullerhb:17}, and is performed by integrating the probability distribution in Figure \ref{KIC81plot} over all $N$ and up to a chosen amplitude $\Delta L/L$. Figure \ref{KIC81numex} shows the results. Our model slightly overpredicts the number of low amplitude TEOs with $\Delta L/L \lesssim 10^{-4}$ but the agreement is fairly good. However, our model indicates the expected number of TEOs with $\Delta L/L \geq 10^{-3}$ is about $0.05$. In other words, only $5 \%$ of systems with properties nearly identical to K81 would be expected to exhibit such a large amplitude TEO. We thus find it unlikely that the primary TEO in K81 is caused by a chance resonance. 

Instead, we suggest that the prominent TEO in K81 is generated via resonance locking (see \citealt{witte:99,witte:01,fullerkoi54:12,burkart:12,burkart:14}). Because the g mode frequencies change as the star evolves, they pass through resonances with tidal forcing frequencies. A resonance lock occurs when tidal dissipation from a resonant mode causes orbital decay such that the tidal forcing frequency increases at the same rate as the mode frequency, and the mode remains resonant. Resonance locking configurations can be stable and last for long periods of time \citep{burkart:14}, and they would create a single high-amplitude TEO like that in K81.

To test the resonance locking hypothesis, we generate stellar models slightly younger and older than our model for K81, and compute their g mode spectra. We then compute the rates at which the mode frequencies evolve, $t_\alpha = \sigma_\alpha/\dot{\sigma}_\alpha$, where $\sigma_\alpha = \omega_\alpha + m \Omega_s$ is the mode frequency in the inertial frame. We assume no angular momentum loss and rigid stellar rotation. In contrast to previous assumptions \citep{fullerkoi54:12,burkart:12}, g mode frequencies typically increase with age in intermediate-mass stars because they become more highly stratified as they evolve. This creates increasing Brunt-V\"{a}is\"{a}l\"{a} frequencies and g mode frequencies, allowing resonance locking to occur with modes of any value of $m$. 

After calculating the mode frequency evolution rates, we calculate corresponding resonance locking mode amplitudes and luminosity fluctuations, 
\beq
\label{reslum}
A_{N,{\rm ResLock}} = \bigg[ \frac{c_\alpha}{\gamma_\alpha t_{\alpha}} \bigg]^{1/2} V_{lm} L_\alpha \, .
\eeq
Here, $c_\alpha$ is a dimensionless factor of order $10^{-2}-10^{-3}$ in our models which is derived in \cite{fullerhb:17}. We plot these predictions for $m=1$ and $m=2$ modes in Figure \ref{KIC81plot}. We have added a $\sim 3 \%$ uncertainty in frequency and $\sim 30 \%$ uncertainty in luminosity fluctuation to account for the uncertainty in the stellar/orbital parameters. The primary TEO has an amplitude and frequency consistent with being a resonantly locked $m=1$ oscillation mode. A resonantly locked $m=2$ mode may be possible but the predicted amplitude is slightly too low. Addtionally, using the system parameters, we evaluate equation 53 of \cite{burkart:14} to find that resonance locks will be stable for both $m=1$ and $m=2$ modes.\footnote{Equation 53 of \cite{burkart:14} has a typo, the $>$ sign should be a $<$ sign.}

In the resonance locking scenario, both the resonantly locked mode frequency $\sigma_\alpha$ and tidal forcing frequency $N \Omega$ are increasing at the same rate, set by the mode frequency evolution time scale $t_\alpha \sim 4 \, {\rm Gyr}$ for $m=1$ g modes in K81. We compute a corresponding tidal orbital evolution time scale of $t_{\rm orb,tide} = E_{\rm orb}/\dot{E}_{\rm orb,tide} \sim 6 \,{\rm Gyr}$, which is somewhat longer than the stellar age in this particular case. This tidal dissipation rate can be translated into an effective tidal quality factor Q \citep{goldreich:66} by defining
\beq
\label{Qtide}
\frac{\dot{E}_{\rm orb,tide}}{E_{\rm orb}} = \frac{3 k_2}{Q} \frac{M'}{M} \bigg(\frac{R}{a_{\rm peri}} \bigg)^5 \Omega \, ,
\eeq
where $a_{\rm peri} = a(1-e)$ is the periastron orbital separation and $k_2 \simeq 3 \! \times \! 10^{-3}$ is the Love number for our model. We calculate $Q \sim 5 \! \times \! 10^4$, much lower than might be naively expected for an F-type star. Indeed, in our model without a resonantly locked mode, the combined energy dissipation of all TEOs leads to $Q \sim 2.5 \! \times \! 10^7$, meaning the resonantly locked mode increases the tidal energy dissipation rate by a factor of $\sim 500$.

\section{Discussion and Conclusions}
\label{disc}

We have demonstrated that resonance locking in KIC8164262 (K81) is an appealing mechanism to account for its prominent tidally excited oscillation (TEO). Resonance locking correctly predicts the amplitude of the TEO, whereas producing the TEO by a chance resonance between tidal forcing and a stellar oscillation mode is unlikely. However, we note that K81 was chosen for detailed analysis because of its high-amplitude TEO, which is larger amplitude than nearly all other TEOs in catalogued heartbeat systems. Of the $\sim \! 175$ catalogued heartbeat stars\footnote{ A list of catalogued heartbeat stars can be found at http://keplerebs.villanova.edu/search, by filtering with the ``HB" flag checked.}, roughly 30 show evidence for TEOs visible by eye. Thus, the chances of finding one system amongst these with an unexpectedly large TEO (at the $2 \sigma$ level) is of order unity. We therefore cannot exclude the possibility the prominent TEO in K81 is simply an uncommon occurrence that was selected for study due to its high amplitude.

K81 can be compared with the KOI-54 \citep{welsh:11} heartbeat system.  Resonance locking with $m=2$ modes was proposed by \cite{fullerkoi54:12} to explain KOI-54's large amplitude TEOs, but \cite{oleary:14} demonstrated that the TEO phases identifies them as $m\!=\!0$ modes. 
%\!\!\footnote{P-mode frequencies do decrease as stars evolve off the main sequence. The result of increasing g-mode and decreasing p-mode frequencies is that the frequency ranges of p-modes and g-modes in evolved stars (sub-giants and red-giants) overlap with one another, i.e., these stars exhibit mixed modes.}
However, g mode frequencies increase as stars evolve off the main sequence and allows resonance locking to occur with $m\!=\!0$ modes, and hence the prominent TEOs in KOI-54 could still be explained as a resonantly locked $m\!=\!0$ TEO within each star. A preliminary calculation indicates this may be possible, but the observed TEO amplitudes are a factor of a few smaller than expected for resonance locking.

%However, the non-linear mode coupling observed in KOI-54 \citep{oleary:14} may greatly increase mode damping rates from the value we calculate via radiative diffusion. Since a resonantly locked mode amplitude scales as $\Delta L/L \propto \gamma_\alpha^{-1/2}$, non-linearly enhanced damping would result in smaller luminosity fluctuations. 

%Both systems exhibit a large amplitude TEO (two in the case of KOI-54) that is difficult to explain by a chance resonance \citep{fullerkoi54:12,burkart:12}. Resonance locking with $m=2$ modes was proposed by \cite{fullerkoi54:12}, but
%However, \cite{fullerkoi54:12} incorrectly assumed that g mode frequencies decrease with time,

Other heartbeat systems may not exhibit resonance locking effects if they are not in an evolutionary stage where resonantly locked modes are visible. Since TEOs are most visible in stars without thick surface convective zones, we only expect to see large amplitude TEOs in stars with $T_{\rm eff} \gtrsim 6500 \,{\rm K}$. Moreover, resonance locking amplitudes increase as the stellar evolution rate accelerates, i.e., when stars evolve off the main sequence. Therefore, we expect to see high-amplitude resonantly locked modes in heartbeat stars containing somewhat massive ($M \! \gtrsim \! 1.5 \, M_\odot$) stars in the brief period during which they are beginning to evolve off the main sequence, but have not yet cooled to $T_{\rm eff} \! \lesssim 6500 \! \,{\rm K}$. A firm conclusion will require detailed analyses of a greater population of heartbeat systems. If resonance locking does commonly occur, it can greatly enhance tidal dissipation,  causing orbital decay and spin synchronization to proceed on a stellar evolution time scale. During a resonance lock, the tidal orbital energy dissipation rate is $\dot{E}_{\rm orb}/E_{\rm orb} \sim 2/(3 t_\alpha)$. 

Resonance locking is not necessarily limited to eccentric binary star systems, and may operate in many astrophysical scenarios, including circular (but non-synchronized) binary stars, exoplanetary systems, inspiraling white dwarfs \citep{burkart:13} and outwardly migrating planetary moon systems \cite{fullersattide:16}. The observable feature of resonance locking is a larger-than-expected variability at an integer multiple of the orbital frequency.\footnote{Resonance locking cannot explain the pulsations in HAT-P-2 reported by \cite{dewit:17} because those pulsations are larger than $A_{N,{\rm max}}$ at the observed pulsation frequency.} This could potentially be detected as ``anomalous" ellipsoidal variations (see e.g. \citealt{borkovits:14}) or a perturbed gravity field. Resonance may not be able to operate in all scenarios, as it could sometimes be quenched by non-linear instabilities or be overwhelmed by other tidal/orbital effects. When resonance locking can operate, it generally accelerates tidal evolution to proceed on the relevant evolutionary timescale, e.g., a stellar evolution timescale, magnetic braking timescale, or gravitational radiation orbital decay timescale.

\section*{Acknowledgments}

We thank the Planet Hunters and Dan Fabrycky for discovering this system, and the anonymous referee for a thoughtful report. JF acknowledges partial support from NSF under grant no. AST-1205732 and through a Lee DuBridge Fellowship at Caltech. KH acknowledges support through NASA ADAP grant (16-ADAP16-0201). This research was supported by the National Science Foundation under Grant No. NSF PHY11-25915, and by NASA under grant 11-KEPLER11-0056.

\bibliography{../Work/references/fuller,../Work/references/heartbeat,../Work/references/neutronstars,../Work/references/massivestars,../Work/references/angmomtrans,../Work/references/astero}

\end{document}